\documentclass[usenatbib]{mn2e}

\usepackage{graphicx}
\usepackage{amssymb}
\usepackage{times}

\def\be{\begin{equation}}
\def\ee{\end{equation}}

\def\bea{\begin{eqnarray}}
\def\eea{\end{eqnarray}}

\def\R{{{\cal{R}}}}

\def\S{{\cal{S}}}

\def\H{{\cal H}}

\def\cs2{c_{\rm{s}}^2}
\def\dPnad{\delta P_{\rm{nad}}}

\def\U0{{\bar U_0}}

\def\bi{\begin{itemize}}
\def\ei{\end{itemize}}

\def\Mpc{{\rm{Mpc}}}

\begin{document}
\title[Magnitude of the non-adiabatic pressure]{The magnitude of the non-adiabatic pressure in the cosmic fluid}
\author[Brown, Christopherson and Malik]{Iain A.~Brown${}^{1}$, Adam J.~Christopherson${}^{2, 3}$ and Karim A.~Malik${}^{3}$ \\
$^1$Institute of Theoretical Astrophysics, University of Oslo, 0315 Oslo, Norway\\
$^2$School of Physics and Astronomy, University of Nottingham, University Park, Nottingham, NG7 2RD,
 United Kingdom\\
$^3$Astronomy Unit, School of Physics and Astronomy, Queen Mary University of London, Mile End Road, London, E1 4NS, United Kingdom
}

\date{\today}

\maketitle

\begin{abstract}
Understanding the non-adiabatic pressure, or relative
entropy, perturbation is crucial for studies of early-universe
vorticity and Cosmic Microwave Background observations. We calculate
the evolution of the linear non-adiabatic pressure perturbation from
radiation domination to late times, numerically solving the linear
governing equations for a wide range of wavenumbers. Using adiabatic
initial conditions consistent with WMAP seven year data, we find
nevertheless that the non-adiabatic pressure perturbation is non-zero
and grows at early times, peaking around the epoch of matter/radiation
equality and decaying in matter domination. At early times or large
redshifts ($z=10,000$) its power spectrum peaks at a comoving
wavenumber $k\approx 0.2 h/\Mpc$, while at late times ($z=500$) it
peaks at $k\approx 0.02 h/\Mpc$.
\end{abstract}


\section{Introduction}
\noindent Our understanding of the physics of the universe is
improving steadily.  Recent observational successes, in particular
Cosmic Microwave Background (CMB) experiments and Large Scale
Structure (LSS) surveys \citep{WMAP,PLANCK,SDSS}, strongly favour the
cosmological standard model, in which the anisotropies of the CMB and
the LSS are sourced by quantum fluctuations formed during a period of
accelerated expansion in the early universe -- inflation
\citep{LLBook}.

Since the decay of the scalar field(s) driving inflation into the standard
matter fields is not fully understood, the power spectrum of the
fluctuations in the scalar field is mapped onto the spectrum of the
comoving curvature perturbation, $\R$, at horizon crossing. Since $\R$
is conserved on large scales for adiabatic perturbations, the
primordial power spectrum is not affected by the small scale physics,
and can be used to set the initial conditions for Boltzmann codes
\citep{COSMICS,CMBFast,CAMB,CMBEasy,CLASS}. The current data allow for
a small non-adiabatic contribution to the primordial perturbations, only
providing an upper bound \citep{WMAP7}.

This contribution is described by the non-adiabatic pressure, or
entropy, perturbation, $\dPnad$. To date this has been calculated
mainly to get a handle on the evolution of the curvature perturbation
\citep{WMLL}, and hence the primordial power spectrum, and also played
a key role in characterising the initial conditions (``adiabatic''
versus ``isocurvature'' \citep{LLBook}). Observations of the CMB are
fully consistent with purely adiabatic initial conditions
\citep{WMAP7}, although a level of primordial isocurvature can be
allowed (references \citep{Valiviita:2009bp,Mangilli:2010ut,Li:2010yb}
provide an non-exhaustive example of recent studies).

However, after the initial conditions have been imposed, there are no
restrictions on the non-adiabatic contribution. In this letter we
focus on the relative entropy perturbation, which naturally arises in
any multi-component system. This contribution to $\dPnad$ has so far
not been studied in detail in realistic models, and we calculate its
evolution numerically, starting from adiabatic initial conditions in
agreement with the WMAP7 concordance model.

A non-zero entropy perturbation sources the evolution of the curvature
perturbation, which will not be conserved on large scales beyond the
time when the initial conditions are set. In practice, this will
affect any calculation assuming conservation of the curvature
perturbation after $z\sim 100,000$. Furthermore, it has been shown
that the non-adiabatic pressure perturbation can generate vorticity in
the early universe \citep{Christopherson:2009bt}. Although this
vorticity is generated at second order in perturbation theory, it
might act as foreground and contribute to the primordial B-mode
polarisation signal in the CMB on small scales.\\

\section{Definitions}
We consider linear, scalar perturbations to a flat
Friedmann-Robertson-Walker (FRW) spacetime, following \citet{MW04}. The total density and the total pressure including
background and perturbations, $\rho$ and $P$, respectively, are
related to the density and pressure of the component fluids by
$\rho=\sum_\alpha\rho_{\alpha}\,, P=\sum_\alpha P_{\alpha}\,$, where
Greek letters label the individual fluids.

The total pressure perturbation can be split into an adiabatic and
non-adiabatic part 
$
\delta P\equiv \delta P_{\rm nad}+c^2_{\rm s}\delta\rho\,,
$
where $\cs2$ is the adiabatic speed of sound, $c^2_{\rm s}\equiv \dot
P / \dot\rho\,$, and an overdot denotes a derivative with respect to
conformal time, $\eta$.

In the presence of more than one fluid, the total non-adiabatic
pressure perturbation, $\delta P_{\rm nad}$, may be further decomposed
\citep{KS} as
$
\delta P_{\rm nad}\equiv \delta P_{\rm intr}+\delta P_{\rm rel}\,.
$
The first part is the sum of the intrinsic entropy perturbation of each
fluid, which vanishes for barotropic fluids.
We assume here for simplicity that all individual fluids have a constant equation of state and are hence barotropic.

The second part of the non-adiabatic pressure perturbation is due to the {\em relative entropy perturbation} $\cal{S}_{\alpha\beta}$ between different fluids, defined as 
\be
\label{defS}
\S_{\alpha\beta} \equiv
-3\H\left(\frac{\delta\rho_\alpha}{\dot\rho_\alpha} -\frac{\delta\rho_\beta}{\dot\rho_\beta} \right) \,,
\ee
where $\H$ is the conformal Hubble parameter, $\H\equiv\dot{a}/a$. The relative non-adiabatic pressure perturbation between multiple fluids is then 
\bea
\label{deltaPrel}
\delta P_{\rm rel}&=&-\frac{1}{6\H\dot\rho} \sum_{\alpha,\beta}\dot\rho_\alpha\dot\rho_\beta \left(c^2_\alpha-c^2_\beta\right)\S_{\alpha\beta}\\ 
&=&\frac{1}{2\dot{\rho}}\sum_{\alpha,\beta} \left(c_\alpha^2-c_\beta^2\right) \left(\dot{\rho}_\beta\delta\rho_\alpha -\dot{\rho}_\alpha\delta\rho_\beta\right)\,, \nonumber
\eea
with $c^2_{\alpha}\equiv \dot P_{\alpha}/ \dot\rho_{\alpha}$ the adiabatic sound speed of each individual fluid. 
It is straightforward to see that the non-adiabatic pressure
perturbation is invariant under a gauge transformation. This also
implies that if it is non-zero in one gauge, it will be non-zero in
any other gauge; i.e.~it can not be ``gauged away''.

By definition, the relative non-adiabatic pressure perturbation does
not contribute to the individual fluid evolution equations and instead
enters the Einstein equations. It is perhaps in this context that the
non-adiabatic pressure perturbation is most familiar, governing the
evolution of the curvature perturbation on uniform density
hypersurfaces, $\zeta$, on large scales \citep{WMLL}.

Boltzmann codes typically avoid the Einstein equations in which the
non-adiabatic pressure perturbation would enter -- these have instead
been used as consistency checks -- and so the system we are evolving
remains closed at linear order.

However, at higher orders in perturbation theory, knowledge of the
behaviour $\delta P_{\rm rel}$ becomes important. In particular, the
non-adiabatic pressure produced by linear perturbations will have a
significant impact on the curvature perturbation on uniform density
hypersurfaces, $\zeta$, at a non-linear level. A first study in this
direction was undertaken at second order by \citet{M2005}, showing
the dependence of $\zeta_2$ on the relative entropy perturbations. It
has also been shown that the non-adiabatic pressure perturbation can
generate vorticity in the early universe
\citep{Christopherson:2009bt}. Furthermore, if a modification to a
Boltzmann code includes a perturbed total pressure the non-adiabatic
pressure perturbation should be included.\\

\section{Dynamics and initial conditions}
\noindent 
The concordance cosmology
contains five fluids: baryonic matter, cold dark matter,
photons, neutrinos and some form of dark energy. We assume the
neutrinos to be massless. Defining the equation of state parameter
$w_\alpha=P_\alpha/\rho_\alpha$, baryons ($b$) and CDM ($c$) have
$w_b=w_c=c_b^2=c_c^2=0$. Photons ($\gamma$) and neutrinos ($\nu$) are
relativistic species and have
$w_\gamma=w_\nu=c_\gamma^2=c_\nu^2=1/3$. Photons and baryons are
coupled together by Compton scattering, while neutrinos stream
freely. We employ a flat $\Lambda$CDM cosmology consistent with the
WMAP 7-year results \citep{WMAP7} with the parameters
\bea
&\Omega_bh^2=2.253\times 10^{-2}\,, \quad \Omega_ch^2=0.112\,,& \nonumber \\
&\Omega_\Lambda=0.728\,, \quad h=0.704 .&
\eea
Although we employ a cosmological constant, our approach also allows
for a dark energy with $w_\phi\neq 1$ and therefore dark energy
density perturbations. These would contribute to the non-adiabatic
pressure perturbation at late times. Since our focus is on the early
universe where dark energy perturbations are expected to contribute
negligibly, results from $\Lambda$CDM will be very similar to those
from more general models. Of course, formally we should include dark
energy perturbations \citep{Hwang:2009zj, Christopherson2010}, and
these will be investigated in future work.

Since the non-adiabatic pressure perturbation is gauge-invariant it can be
evaluated in any gauge, and we choose for numerical simplicity the  synchronous
gauge, 
in which the line-element takes the form
\be
ds^2=a^2(\eta)\left(-d\eta^2+\left(
\delta_{ij}(1-\psi)+E_{,ij}\right)dx^idx^j\right)\,,
\ee
employing the conformal time $\eta$. The residual gauge freedom is
removed by choosing to co-move with cold dark matter,
i.e.~$v_c(k,\eta)=0$. The governing equations are then presented in
\citet{MaBert}.\footnote{In terms of the variables $h$ and $\eta_{\rm
MB}$ in \citet{MaBert}, the spatial curvature perturbations $\psi$
and $E$ (in \citet{MW04}) are $\psi=\eta_{\rm{MB}}$ and
$E=-(h+6\eta_{\rm MB})/(2k^2)$, while the velocity perturbation is
$v_\alpha=\theta_\alpha/k^2$.}

Taking the entropy perturbation between each species to vanish at some early epoch gives the fluids the initial conditions in tight-coupling and deep radiation domination \citep{MaBert}
\bea
\label{eq:IC1}
&\delta_\gamma=\delta_\nu=\frac{4}{3}\delta_b=\frac{4}{3}\delta_c=-\frac{2}{3}Ck^2\eta_i^2\,,& \nonumber \\
&v_\gamma=v_b=-\frac{1}{18}Ck^2\eta_i^3, \quad v_\nu=\frac{23+4R_\nu}{15+4R_\nu}v_\gamma\,, \quad v_c\equiv 0\, .& 
\eea
These are supplemented by initial conditions for $\psi$, $E$ and the neutrino shear which we do not present here; for clarity we have also neglected small corrections due to the non-zero matter density even in the early times, although they are included in the code. These initial conditions are implemented at $z_i\approx 100,000$ to ensure that all linear modes are strongly superhorizon when the code begins. The constant $R_\nu=\rho_\nu/(\rho_\gamma+\rho_\nu)$ is the relative neutrino abundance.

\begin{figure}
\begin{center}\includegraphics[width=0.6\columnwidth]{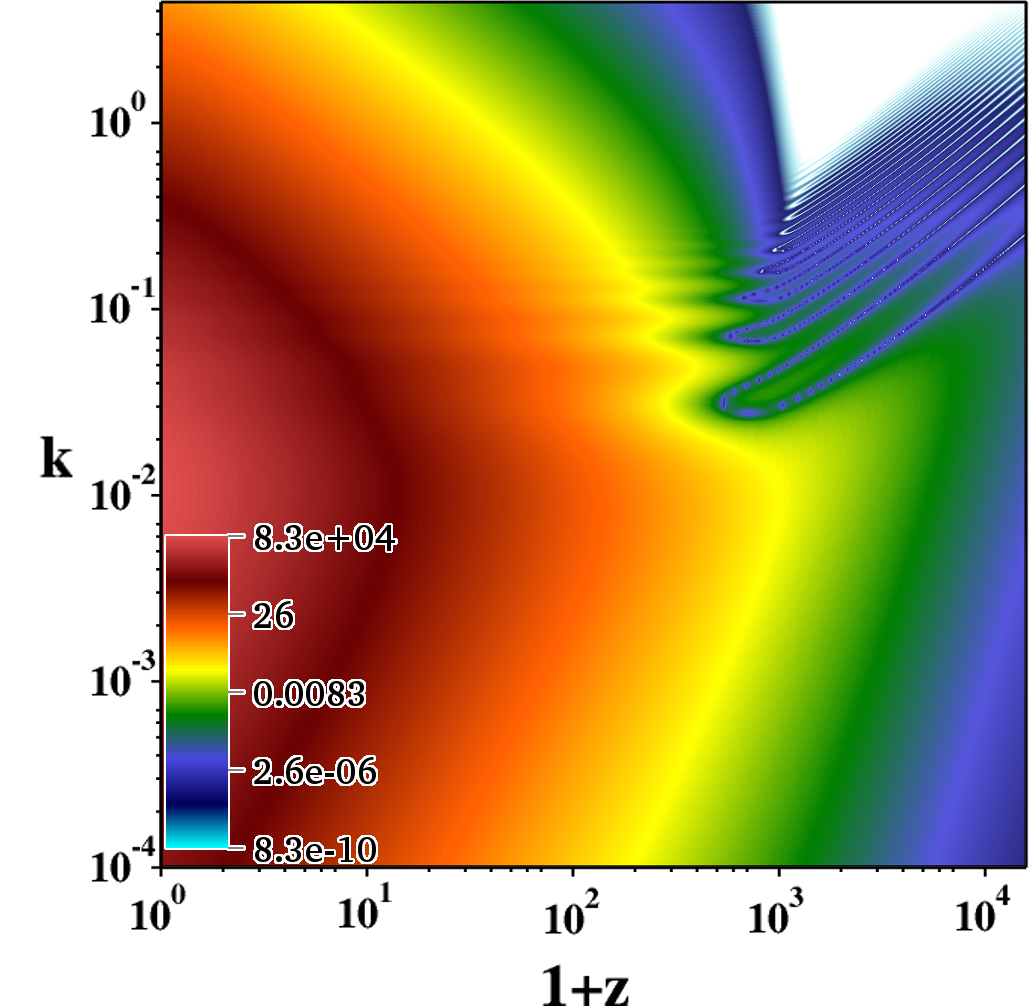}\end{center}
\caption{Power spectrum of the baryon density contrast, $P_b(k,\eta)$.}
\label{BaryonPert}
\end{figure}

The system of equations can now be integrated to find the fluid and metric perturbations at arbitrary time. The power spectrum of a perturbation $\delta(\mathbf{k})$ is given by
\bea
P_\delta(k,\eta)&=&\left<\delta(\mathbf{k})\delta^*(\mathbf{k}')\right>\nonumber \\ &=&\frac{2\pi^2}{k^3}\mathcal{P}_\zeta(k)\left|\delta(k)\right|^2(2\pi)^3\delta^3(\mathbf{k}-\mathbf{k}')
\eea
where
$\mathcal{P}_\zeta(k)=A_\star(k/k_\star)^{n_s-1}$
is defined using the curvature perturbation $\zeta\approx\mathcal{R}$ which is constant on superhorizon scales. The amplitude $A_\star$ at the pivot wavenumber $k_\star$ take the WMAP7 values $A_\star=2.42\times 10^{-9}$ at $k_\star=0.002h\rm{Mpc}^{-1}$. Figure~\ref{BaryonPert} shows the power spectrum of the baryon density contrast $\delta_b(k)$,
\be
P_b(k,\eta)=\frac{2\pi^2}{k^3}\mathcal{P}_\zeta(k)\left|\delta_b(k)\right|^2\,,
\ee
for the range of wavenumbers and redshifts we sample. We note that the maximum plotted redshift $z_{\rm max}\approx 50,000$ and is therefore significantly later than the initial redshift $z_i\approx 100,000$.

\section{Results}
\noindent
We can now derive an analytical approximation for the non-adiabtic
pressure perturbation.  Defining the scaled entropy differences
between two fluids fluids by
\be
\Delta_{\alpha\beta}=(1+w_\beta)\delta_\alpha-(1+w_\alpha)\delta_\beta
\,,
\ee 
and using the fact that the density contrasts evolve as 
\be
\begin{array}{c}
\dot{\delta_c}+\frac{1}{2}\dot{h}=0\,,\quad
\dot{\delta_b}+\frac{1}{2}\dot{h}+k^2v_b=0\,, \\
\dot{\delta_\gamma}+\frac{2}{3}\dot{h}+\frac{4}{3}k^2v_\gamma=0\,,\quad
\dot{\delta_\nu}+\frac{2}{3}\dot{h}+\frac{4}{3}k^2v_\nu=0\,,
\end{array}
\ee
we find that the scaled entropy differences are governed by
\be
\label{DeltaEvolution}
\begin{array}{c}
\dot{\Delta}_{\gamma b}=\frac{4}{3}k^2\left(v_b-v_\gamma\right)\,,\quad
\dot{\Delta}_{\gamma c}=-\frac{4}{3}k^2v_\gamma\,, \\
\dot{\Delta}_{\nu b}=\frac{4}{3}k^2\left(v_b-v_\nu\right)\,,\quad
\dot{\Delta}_{\nu c}=-\frac{4}{3}k^2v_\nu\, .
\end{array}
\ee
These equations are gauge-invariant, and apply at all
times.

From Eq.~(\ref{eq:IC1}) we see that on superhorizon scales, in
radiation domination and during tight-coupling, the CDM terms will
dominate. Furthermore, the present neutrino abundance implies $R_\nu\approx 0.4$ and so
\be
\label{hierarchy}
\Delta_{\nu c}>\Delta_{\gamma c}\gg\Delta_{\nu b}\gg\Delta_{\gamma b} \,.
\ee
Assuming tight-coupling to be exact, implying $\Delta_{\gamma b}=0$, integrating Eqs.~(\ref{DeltaEvolution}) in the early universe then gives
\be
\!\delta P_{\rm rel}\!\approx\!\frac{3/216 \times H_0^2\Omega_c (15+12R_\nu)}{8\pi G\Big(1+\frac{3}{4}a\frac{\Omega_M}{\Omega_R}\Big)\!(15+4R_\nu)}Ck^4\eta^4a^{-3} 
\!\propto \!k^4a\,.\!\!\!\!\!
\ee
In finding this we have used that in radiation domination $a\propto\eta$. A rough approximation to the behaviour in early matter domination can be found by setting $a\propto\eta^2$, giving the rapidly decaying
\be
\label{MatterDomApprox}
\delta P_{\rm rel}\propto k^4a^{-2}(1+3\Omega_Ma/4\Omega_R)^{-1}
\ee
which suggests that $\delta P_{\rm rel}$ reaches a maximum at or
around the epoch of matter/radiation equality. However, 
Eq.~(\ref{MatterDomApprox}) applies only very close to equality as it
assumes a similar growth of perturbations to that in radiation
equality; the most we can say is that we expect the non-adiabatic
pressure perturbation to decay early in matter domination. This will
be confirmed by a more accurate, numerical analysis.

Given the primordial power spectrum $\mathcal{P}_\zeta(k)\propto k^{n_S-1}$ the power spectrum of the non-adiabatic pressure perturbation in the early universe is then
\be
\left<\delta P_{\rm rel}(\mathbf{k})\delta P^*_{\rm rel}(\mathbf{k}')\right>\propto  k^{4+n_S}a^2 .
\ee
The non-adiabatic pressure perturbation on large scales and at early times grows with scale factor. It is also strongly dependent on $k$ and decreases rapidly on superhorizon scales, which is to be expected.

To calculate the evolution of the non-adiabatic pressure
perturbation we employ a modified version of the CMBFast code
\citep{CMBFast,COSMICS} to recover the velocity perturbations for the
full system from radiation domination to the present day, across a
wide range of scales. The non-adiabatic pressure perturbation is then
found by integrating Eqs.~(\ref{DeltaEvolution}).

\begin{figure}[h]
\begin{center}
\includegraphics[width=0.6\columnwidth]{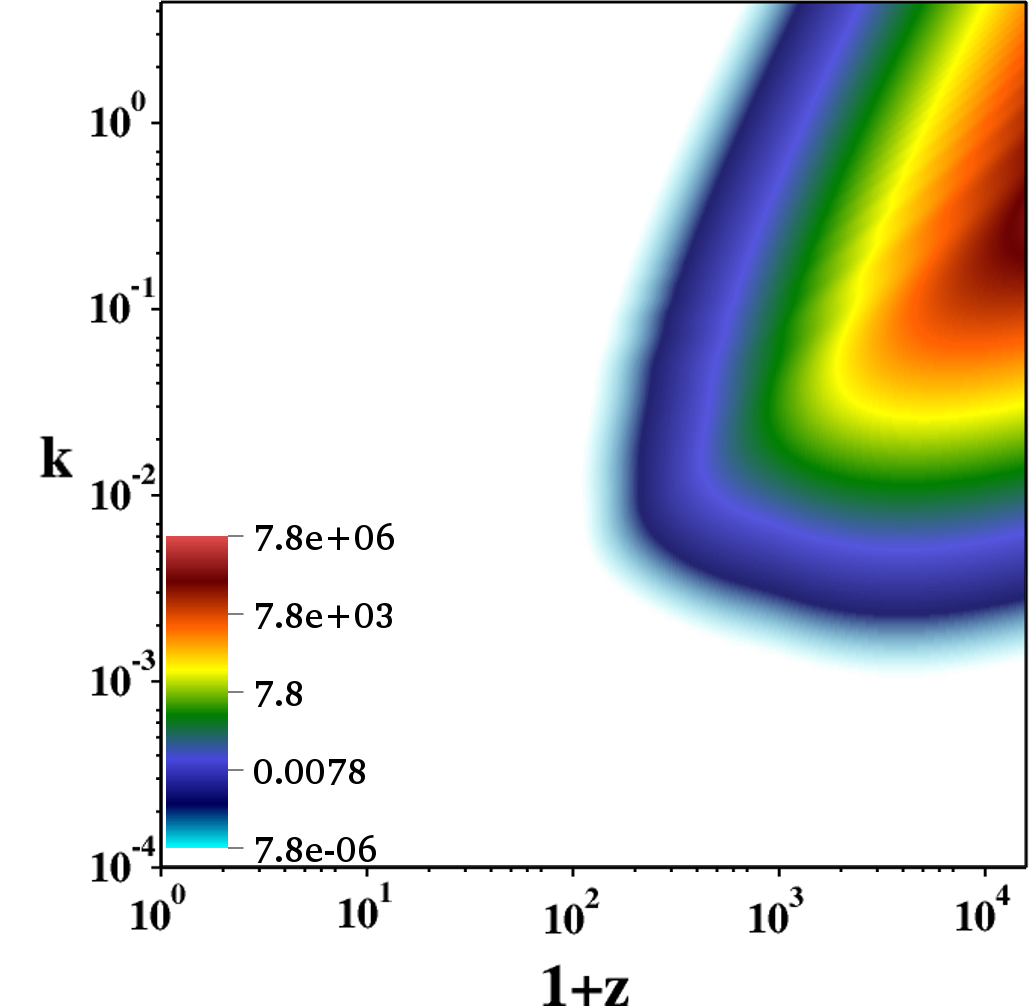}\end{center}
\caption{Power spectrum of the non-adiabatic pressure perturbation $P_{\delta P_{\rm rel}(k,\eta)}$.}
\label{DeltaPRel}
\end{figure}

The power spectrum of the non-adiabatic pressure perturbation is plotted across scale and redshift in Figure \ref{DeltaPRel}. The oscillations before recombination are prominent, and $\delta P_{\rm rel}$ indeed peaks at approximately matter/radiation equality. It is orders of magnitude more significant on smaller scales than on larger scales and at early times peaks at approximately $k\sim 0.1h/\rm{Mpc}$. Across all wavenumbers $\delta P_{\rm rel}$ decays rapidly after matter/radiation equality. The non-adiabatic pressure perturbation is negative.

\begin{figure*}
\begin{center}
\includegraphics[width=0.6\textwidth]{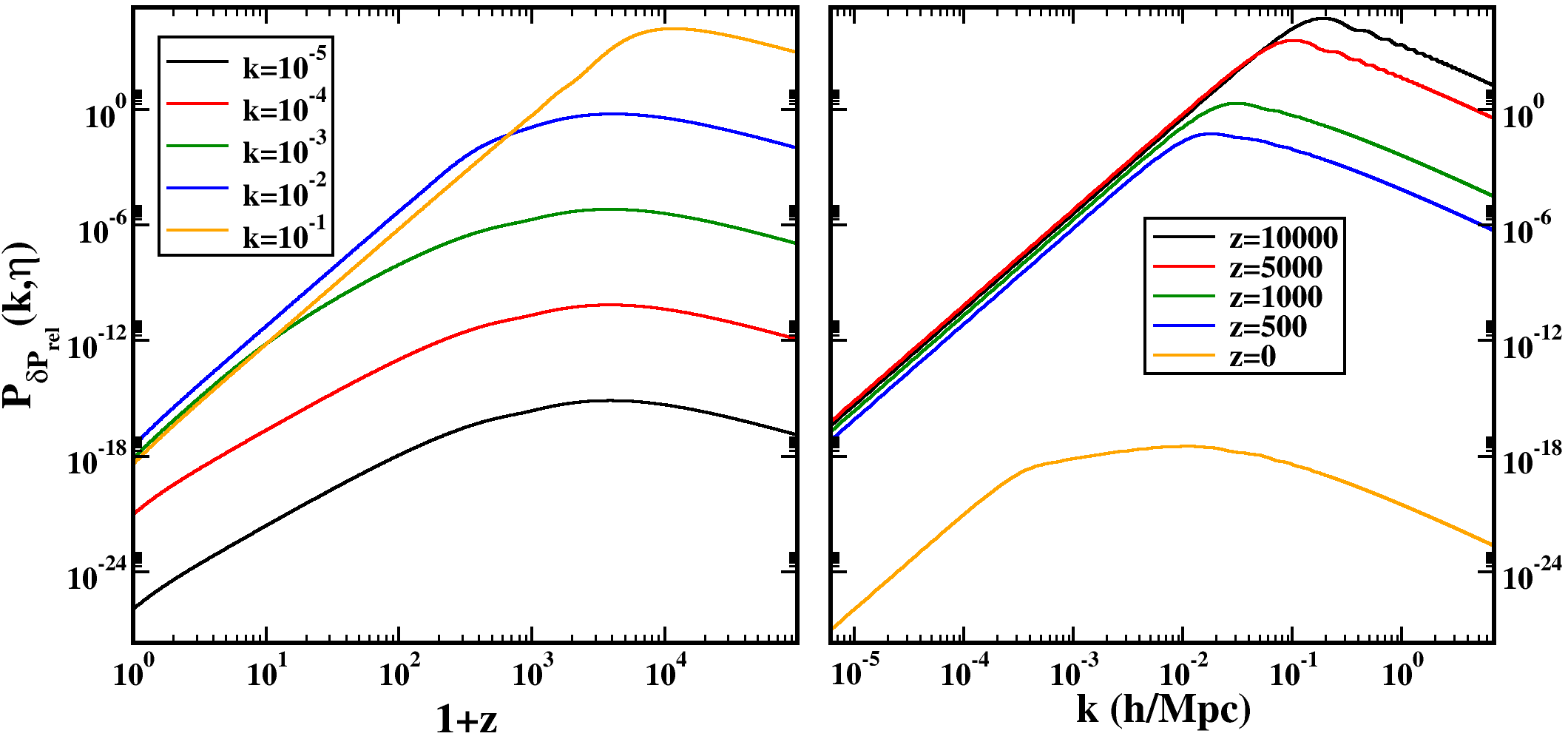}\end{center}
\caption{Power spectrum of the non-adiabatic pressure perturbation $P_{\delta P_{\rm rel}(k,\eta)}$ as a function of redshift for set wavenumber (left); and as a function of wavenumber for set redshift (right).}
\label{DeltaPRelKZ}
\end{figure*}

The left panel of Figure~\ref{DeltaPRelKZ} shows the evolution of this non-adiabatic pressure perturbation for wavelengths between $k=10^{-5}h\rm{Mpc}^{-1}$ and $k=10^{-1}h\rm{Mpc}$. The growth with scale-factor at the earliest times is apparent, as is the decay during matter domination. However, the period at which $\delta P_{\rm rel}$ peaks is extended, being approximately within an order-of-magnitude of the peak between $z\approx 10,000$, significantly before matter/radiation equality, and recombination at $z\approx 1,100$, after which it decays rapidly. For very short wavelengths $\delta P_{\rm rel}$ peaks at earlier times and decays more rapidly. The decay after recombination goes as $P_{\delta P_{\rm rel}}\propto a^{-6}$ for smaller scales and as $P_{\delta P_{\rm rel}}\propto a^{-5}$ for larger scales, implying $\delta P_{\rm rel}\propto a^{-3}$ and $\delta P_{\rm rel}\propto a^{-5/2}$ on small and large scales respectively.

The right panel of Figure~\ref{DeltaPRelKZ} shows the dependence of the non-adiabatic pressure perturbation on wavenumber for a range of redshifts from deep radiation domination to the present epoch. The decay towards the present epoch is clear, and the scale at which the perturbation has most power grows from $k\sim 0.2$ in deep radiation domination to $k\sim 0.02$ at a redshift of $z\approx 500$. The power spectrum scales as $P_{\delta P_{\rm rel}}\propto k^{-5/2}$ on small scales and as $P_{\delta P_{\rm rel}}\propto k^5$ on large scales. The non-adiabatic pressure perturbation itself then scales as $\delta P_{\rm rel}\propto k^{1/4}$ on small scales and $\delta P_{\rm rel}\propto k^4$ on large scales.

These results can be summarised as
\be
\delta P_{\rm rel}\propto a\left\{\begin{array}{rl} k^4,&k\lesssim 10^{-1}h/\Mpc \\ k^{1/4},& k\gtrsim 10^{-1}h/\Mpc\end{array}\right.
\ee
at early times, and
\be
\delta P_{\rm rel}\propto \left\{\begin{array}{rl} a^{-7/2}k^{4},&k\lesssim 10^{-3}-10^{-4}h/\Mpc \\ a^{-3}k^{1/3},& k\gtrsim 10^{-1}h/\Mpc\end{array} \right.
\ee
at late times. This contrasts with \citet{Christopherson:2010ek},
which assumed $\delta P_{\rm nad}\propto k^\alpha a^{-5}$ with
$\alpha\neq 1$ before setting $\alpha=2$. This choice underestimates
the gradient and overestimates the rate of decay of $\delta P_{\rm
nad}$ on large scales while failing to model the behaviour on small
scales. We would like to emphasise that we are working with small but finite wavenumbers; in the limit $k\rightarrow 0$ the non-adiabaticity vanishes, and the rapid decay ($\sim k^4$) on superhorizon scales is very clear.

\begin{figure*}
\begin{center}
\includegraphics[width=0.6\textwidth]{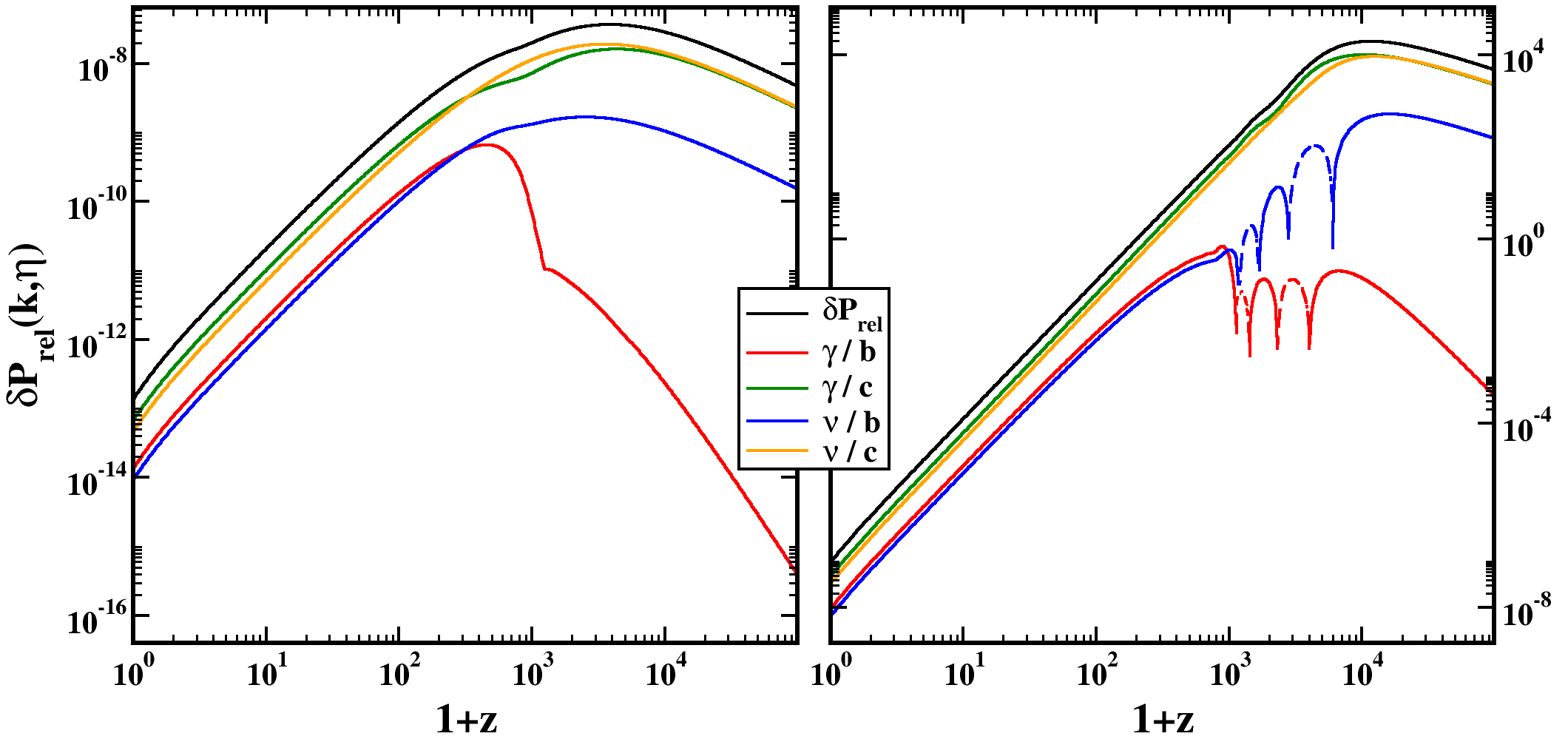}
\end{center}
\caption{The contributions to the total relative non-adiabatic pressure perturbation arising from the photon/baryon, photon/CDM, neutrino/baryon and neutrino/CDM couplings at $k=10^{-4}h\rm{Mpc}^{-1}$ (left) and $k=10^{-1}h\rm{Mpc}^{-1}$ (right). Absolute values are plotted; the terms are all negative at late epochs.}
\label{RelativeImportance}
\end{figure*}

The contribution to the non-adiabatic pressure perturbation due to the entropy difference between fluid species $\alpha$ and $\beta$ is
\be
\delta P_{{\rm rel},\alpha\beta}=\frac{\rho_\alpha\rho_\beta\left((1+w_\beta)\delta_\alpha-(1+w_\alpha)\delta_\beta\right)}{3(\rho+p)} \,.
\ee
Figure~\ref{RelativeImportance} shows the contributions of these terms at wavenumbers $k=10^{-4}h\rm{Mpc}^{-4}$ and $k=10^{-1}h\rm{Mpc}^{-1}$. At early times, the CDM/neutrino contribution dominates slightly over the CDM/photon contribution, as was expected in Eq.~(\ref{hierarchy}). The baryon/neutrino term is significantly smaller, and the baryon/photon term (as expected) is smaller again.

After recombination, this picture changes. The decoupling of the photons and baryons causes their relative entropy to increase significantly until at $z\approx 100$, $\delta P_{{\rm rel},\gamma b}\gtrsim \delta P_{{\rm rel},\nu b}$. At late times the baryon perturbations and the CDM perturbations become approximately equal, while $v_\nu$ and $v_\gamma$ behave similarly to one-another. It is no surprise that the CDM/neutrino and CDM/photon contributions exhibit the same behaviour as the baryon/neutrino and baryon/photon terms, and so after recombination, $\delta P_{{\rm rel},\gamma c}>\delta P_{{\rm rel},\nu c}$. The difference between the CDM/radiation and baryon/radiation terms is then due to the relative abundance of CDM over baryons; in the concordance model, the CDM terms dominate by a factor of 4 over the baryon terms.

On small scales the behaviour is similar to the large-scale behaviour,
except that $\delta P_{{\rm rel},\gamma c}\gtrsim\delta P_{{\rm rel},\nu
c}$ for the entire period sampled. For $z\gtrsim 5\times 10^4$ this
would be expected to be reversed. For all times and across all linear
scales, one can therefore find a reasonable approximation to the
non-adiabatic density perturbation by considering only the relative
entropies between the radiative species and the cold dark matter.

\section{Discussion}
\noindent
We have calculated the non-adiabatic pressure perturbation in the
WMAP7 concordance cosmology, finding both its magnitude and its
scaling properties. It is non-vanishing and grows in the early
universe, becoming most significant at around matter/radiation
equality. For a wide range of wavenumbers it remains within an order
of magnitude of its peak until recombination after which it decays
rapidly.

Although the existence of the non-adiabatic pressure perturbation does
not influence existing \emph{first-order} CMB calculations employing
Boltzmann codes such as CAMB or CMBFast, its study is nevertheless extremely
important. Firstly, this quantity is not actually calculated in
Boltzmann codes and yet has a concrete physical meaning -- it is
directly related to the entropy perturbation. One example we used
to motivate its importance is the vorticity it will produce at
second-order, the study of which is ongoing, and the magnitude of
which could be sufficient as to provide an additional foreground to
the CMB polarisation signals. In turn this may produce magnetic fields
that could influence recombination physics or provide seed fields for
the observed cluster magnetic fields, or to influence large-scale
structure formation in the more recent universe. Secondly, even though
the non-adiabatic pressure is formed of linear combinations of
quantities evaluated in Boltzmann codes, it cannot be accurately
recovered by naively combining these quantities. We have assumed
entirely adiabatic initial conditions. On the largest scales and at
early times, then, we have for instance
$\delta_\gamma\approx (4/3)\delta_b$. The non-adiabatic pressure depends on
the difference between these two quantities, which is extremely
small. Calculating this numerically is then rather unstable; naively
reconstructing the non-adiabatic pressure simply from the
perturbations in one of the Boltzmann codes could give entirely the
wrong behaviour on very large scales and at very early
times.\footnote{We found an extreme case where the reconstructed non-adiabatic
pressure \emph{decays} at early times, since the rounding error in
$\delta_\gamma-(4/3)\delta_b$ grows ever more dominant for increasing
redshifts.} As such, one needs to modify the codes to evaluate the
entropy differences directly, as we have in this work.

This perturbation will have various effects. Whereas the curvature
perturbation is conserved on large scales at early times while the
universe can be modelled by a single barotropic fluid, this is no
longer case at later times when the universe has to be modelled as a
multi-fluid system. Another effect is the generation of vorticity at second order. The
assumptions in \citet{Christopherson:2010ek} were conservative;
with the scalings found in our numerical study, we might expect the
resulting vorticity to be enhanced. This will form the focus of a
future study. 
In summary, we have shown that the non-adiabatic pressure perturbation
is non-zero in the early universe for generic initial conditions and
matter content, even if the intrinsic entropy perturbations are
negligible.

{\bf Acknowledgements} The authors are grateful to Ellie Nalson for useful discussions.
IAB would like to thank Frode Hansen and Jussi
V\"alliviita for helpful discussions, and the Astronomy Unit at QMUL
for hospitality. AJC is funded by the Sir Norman Lockyer fellowship of the Royal Astronomical Society. 
KAM is supported in part by the STFC under Grants
ST/G002150/1 and ST/H002855/1.

\bibliographystyle{mn2e}
\bibliography{bcm}

\begin{thebibliography}{}

\bibitem[\protect\citeauthoryear{Abazajian et~al.,}{Abazajian
  et~al.}{2009}]{SDSS}
Abazajian K.~N.,  et~al., 2009, Astrophys. J. Suppl., 182, 543

\bibitem[\protect\citeauthoryear{Ade et~al.,}{Ade  et~al.}{2011}]{PLANCK}
Ade P. A.~R.,  et~al., 2011, Astron. Astrophys., 536, A1

\bibitem[\protect\citeauthoryear{Bertschinger}{Bertschinger}{1995}]{COSMICS}
Bertschinger E.,  1995

\bibitem[\protect\citeauthoryear{Christopherson}{Christopherson}{2010}]{Christopherson2010}
Christopherson A.~J.,  2010, Phys. Rev., D82, 083515

\bibitem[\protect\citeauthoryear{Christopherson, Malik \&
  Matravers}{Christopherson et~al.}{2009}]{Christopherson:2009bt}
Christopherson A.~J.,  Malik K.~A.,    Matravers D.~R.,  2009, Phys. Rev., D79,
  123523

\bibitem[\protect\citeauthoryear{Christopherson, Malik \&
  Matravers}{Christopherson et~al.}{2011}]{Christopherson:2010ek}
Christopherson A.~J.,  Malik K.~A.,    Matravers D.~R.,  2011, Phys. Rev., D83,
  123512

\bibitem[\protect\citeauthoryear{Doran}{Doran}{2005}]{CMBEasy}
Doran M.,  2005, JCAP, 0510, 011

\bibitem[\protect\citeauthoryear{Kodama \& Sasaki}{Kodama \& Sasaki}{1984}]{KS}
Kodama H.,  Sasaki M.,  1984, Prog. Theor. Phys. Suppl., 78, 1

\bibitem[\protect\citeauthoryear{Komatsu et~al.,}{Komatsu
  et~al.}{2011a}]{WMAP}
Komatsu E.,  et~al., 2011a, Astrophys.J.Suppl., 192, 18

\bibitem[\protect\citeauthoryear{Komatsu et~al.,}{Komatsu
  et~al.}{2011b}]{WMAP7}
Komatsu E.,  et~al., 2011b, Astrophys.J.Suppl., 192, 18

\bibitem[\protect\citeauthoryear{Lesgourgues}{Lesgourgues}{2011}]{CLASS}
Lesgourgues J.,  2011

\bibitem[\protect\citeauthoryear{Lewis, Challinor \& Lasenby}{Lewis
  et~al.}{2000}]{CAMB}
Lewis A.,  Challinor A.,    Lasenby A.,  2000, Astrophys. J., 538, 473

\bibitem[\protect\citeauthoryear{Li, Liu, Xia \& Cai}{Li
  et~al.}{2011}]{Li:2010yb}
Li H.,  Liu J.,  Xia J.-Q.,    Cai Y.-F.,  2011, Phys. Rev., D83, 123517

\bibitem[\protect\citeauthoryear{Liddle \& Lyth}{Liddle \& Lyth}{2000}]{LLBook}
Liddle A.~R.,  Lyth D.~H.,  2000, Cosmological inflation and large-scale
  structure.
CUP, Cambridge, UK

\bibitem[\protect\citeauthoryear{Ma \& Bertschinger}{Ma \&
  Bertschinger}{1995}]{MaBert}
Ma C.-P.,  Bertschinger E.,  1995, Astrophys.J., 455, 7

\bibitem[\protect\citeauthoryear{Malik}{Malik}{2005}]{M2005}
Malik K.~A.,  2005, JCAP, 0511, 005

\bibitem[\protect\citeauthoryear{Malik \& Wands}{Malik \& Wands}{2005}]{MW04}
Malik K.~A.,  Wands D.,  2005, JCAP, 0502, 007

\bibitem[\protect\citeauthoryear{Mangilli, Verde \& Beltran}{Mangilli
  et~al.}{2010}]{Mangilli:2010ut}
Mangilli A.,  Verde L.,    Beltran M.,  2010, JCAP, 1010, 009

\bibitem[\protect\citeauthoryear{Park, Hwang, Lee \& Noh}{Park
  et~al.}{2009}]{Hwang:2009zj}
Park C.-G.,  Hwang J.-c.,  Lee J.-h.,    Noh H.,  2009, Phys. Rev. Lett., 103,
  151303

\bibitem[\protect\citeauthoryear{Seljak \& Zaldarriaga}{Seljak \&
  Zaldarriaga}{1996}]{CMBFast}
Seljak U.,  Zaldarriaga M.,  1996, Astrophys. J., 469, 437

\bibitem[\protect\citeauthoryear{Valiviita \& Giannantonio}{Valiviita \&
  Giannantonio}{2009}]{Valiviita:2009bp}
Valiviita J.,  Giannantonio T.,  2009, Phys. Rev., D80, 123516

\bibitem[\protect\citeauthoryear{Wands, Malik, Lyth \& Liddle}{Wands
  et~al.}{2000}]{WMLL}
Wands D.,  Malik K.~A.,  Lyth D.~H.,    Liddle A.~R.,  2000, Phys. Rev., D62,
  043527

\end{thebibliography}

\end{document}